\documentclass[twocolumn,showpacs,amsmath,amssymb,prl,superscriptaddress,10pt]{revtex4-1}

\usepackage{lineno}
\usepackage{graphicx}
\usepackage{amsmath}
\usepackage{hyperref}
\usepackage{braket}
\usepackage{dcolumn}
\usepackage{bm}
\usepackage{upgreek}
\usepackage{threeparttable}
\usepackage{xcolor}

\begin{document}



\title{Ground State Electromagnetic Moments of $^{37}$Ca}



\author{A. Klose}
\affiliation{Department of Chemistry, Augustana University, Sioux Falls, South Dakota 57197, USA}

\author{K. Minamisono}
\affiliation{National Superconducting Cyclotron Laboratory, Michigan State University, East Lansing, MI 48824, USA}
\affiliation{Department of Physics and Astronomy, Michigan State University, East Lansing, MI 48824, USA}

\author{A. J. Miller}
\affiliation{National Superconducting Cyclotron Laboratory, Michigan State University, East Lansing, MI 48824, USA}
\affiliation{Department of Physics and Astronomy, Michigan State University, East Lansing, MI 48824, USA}

\author{B. A. Brown}
\affiliation{National Superconducting Cyclotron Laboratory, Michigan State University, East Lansing, MI 48824, USA}
\affiliation{Department of Physics and Astronomy, Michigan State University, East Lansing, MI 48824, USA}

\author{D. Garand}
\affiliation{National Superconducting Cyclotron Laboratory, Michigan State University, East Lansing, MI 48824, USA}

\author{J. D. Holt}
\affiliation{TRIUMF, Vancouver, BC V6T 2A3, Canada}

\author{J. D. Lantis}
\affiliation{National Superconducting Cyclotron Laboratory, Michigan State University, East Lansing, MI 48824, USA}
\affiliation{Department of Chemistry, Michigan State University, East Lansing, MI 48824, USA}

\author{Y. Liu}
\affiliation{Physics Division, Oak Ridge National Laboratory, Oak Ridge, TN 37831, USA}

\author{B. Maa$\ss$}
\affiliation{Institut f{\"u}r Kernphysik, Technische Universit{\"a}t Darmstadt, 64289 Darmstadt, Germany}

\author{W. N{\"o}rtersh{\"a}user}
\affiliation{Institut f{\"u}r Kernphysik, Technische Universit{\"a}t Darmstadt, 64289 Darmstadt, Germany}

\author{S. V. Pineda}
\affiliation{National Superconducting Cyclotron Laboratory, Michigan State University, East Lansing, MI 48824, USA}
\affiliation{Department of Chemistry, Michigan State University, East Lansing, MI 48824, USA}

\author{D. M. Rossi}
\affiliation{Institut f{\"u}r Kernphysik, Technische Universit{\"a}t Darmstadt, 64289 Darmstadt, Germany}

\author{A. Schwenk}
\affiliation{Institut f{\"u}r Kernphysik, Technische Universit{\"a}t Darmstadt, 64289 Darmstadt, Germany}
\affiliation{ExtreMe Matter Institute EMMI, GSI Helmholtzzentrum f\"ur Schwerionenforschung GmbH, 64291 Darmstadt, Germany}
\affiliation{Max-Planck-Institut f\"ur Kernphysik, Saupfercheckweg 1, 69117 Heidelberg, Germany}

\author{F. Sommer}
\affiliation{Institut f{\"u}r Kernphysik, Technische Universit{\"a}t Darmstadt, 64289 Darmstadt, Germany}

\author{C. Sumithrarachchi}
\affiliation{National Superconducting Cyclotron Laboratory, Michigan State University, East Lansing, MI 48824, USA}

\author{A. Teigelh\"ofer}
\affiliation{TRIUMF, Vancouver, BC V6T 2A3, Canada}

\author{J. Watkins}
\affiliation{National Superconducting Cyclotron Laboratory, Michigan State University, East Lansing, MI 48824, USA}
\affiliation{Department of Physics and Astronomy, Michigan State University, East Lansing, MI 48824, USA}


\date{\today}

\begin{abstract}
The hyperfine coupling constants of neutron deficient $^{37}$Ca were deduced from the atomic hyperfine spectrum of the $4s~^2S_{1/2}$ $\leftrightarrow$ $4p~^2P_{3/2}$ transition in Ca II, measured using the collinear laser spectroscopy technique. 
The ground-state magnetic-dipole and spectroscopic electric-quadrupole moments were determined for the first time as $\mu = +0.7453(72) \mu_N$ and $Q = -15(11)$ $e^2$fm$^2$, respectively. 
The experimental values agree well with nuclear shell model calculations using the universal sd model-space Hamiltonians versions A and B (USDA/B) in the $sd$-model space with a 95\% probability of the canonical nucleon configuration. 
It is shown that the magnetic moment of $^{39}$Ca requires a larger non-$sd$-shell component than that of $^{37}$Ca for good agreement with the shell-model calculation, indicating a more robust closed sub-shell structure of $^{36}$Ca at the neutron number $N$ = 16 than $^{40}$Ca. The results are also compared to valence-space in-medium similarity renormalization group calculations based on chiral two- and three-nucleon interactions. 
\end{abstract}


\maketitle

{\it Introduction} --- 

A nucleus with finite spin possesses electromagnetic moments capable of providing critical information for the investigation of nuclear structure. Most notably, close to nuclei with magic numbers of nucleons (e.g. 2, 8, 20, 28, 50, 82...) such systems, due to their simple and robust structures, provide discerning comparisons between experiment and theory. One of the highlights of modern nuclear structure studies has been the disappearance of established magic numbers (essentially those seen in stable nuclei \cite{may49,hax49}) and the appearance of new magic numbers at extreme neutron-to-proton ratios \cite{oza00}, for example the neutron number $N$ 
= 16 \cite{kan09,hof09}. 


The ground-state electromagnetic moment of $^{37}$Ca, which has one neutron added to the $N$ =16 $^{36}$Ca nucleus in the vicinity of the proton dripline, was determined in the present study.
The neutron occupies the $d_{3/2}$ orbital with $ j_{<} \equiv l - 1/2$ where $l$ is the orbital angular momentum. The spin-orbit partner $d_{5/2}$ with $ j_{>} \equiv l + 1/2$ is fully occupied. Here, the first-order core-polarization effect, first introduced by Arima and Horie \cite{ari54a,ari54b}, is expected to play an important role on the magnetic moment due to collective M1 excitations between the spin-orbit partners. The counterpart is a $^{39}$Ca nucleus with one neutron hole in the doubly-closed $^{40}$Ca core in the $  j_{<} \equiv l - 1/2  $ shell. In the single-particle model magnetic moments of  $^{37}$Ca and $^{39}$Ca are equal, and take the Schmidt value. This provides a unique situation where the transition from the $JJ$ closed sub-shell $^{36}$Ca to the $LS$ doubly closed $^{40}$Ca configuration can be seen in nearby isotopes in a single element. The variation of structure around the $^{36}$Ca and $^{40}$Ca nuclei is investigated through the first order core-polarization model in the context of the universal sd model-space Hamiltonians versions A and B (USDA/B) and the chiral effective field theory. 

{\it Experiment} --- 
The radioactive ion beam of $^{37}$Ca ($I^\pi$ = 3/2$^+$, $T_{1/2}$ = 175 ms) was produced via projectile-fragmentation reactions of a $^{40}$Ca primary beam on a Be target. The $^{40}$Ca beam was accelerated to 140 MeV/nucleon in the coupled cyclotrons at the National Superconducting Cyclotron Laboratory at Michigan State University. The $^{37}$Ca fragments were separated using the A1900 fragment separator \cite{mor03}, thermalized in a He-filled gas cell \cite{coo14} and extracted as singly-charged ions at an energy of 30 keV. The low-energy beam was then mass analyzed through a dipole magnet and transported to the BEam COoling and LAser spectroscopy (BECOLA) facility \cite{min13,ros14}. The typical rate of $^{37}$Ca at the BECOLA facility was 10$^3$ ions/s. 

At BECOLA, the $^{37}$Ca beam was first injected into a radio frequency quadrupole (RFQ) cooler and buncher \cite{bar17}. The ion beam was trapped, cooled (improving the emittance) and extracted at an approximate energy of 29850~eV as ion bunches for the bunched-beam collinear laser spectroscopy \cite{cam02,nie02}. Laser-induced fluorescence measurements were performed on the $4s~^2S_{1/2}$ $\leftrightarrow$ $4p~^2P_{3/2}$ transition in Ca II at 393~nm to measure the hyperfine (hf) spectrum. The ion-beam bunch was extracted from the RFQ every 330~ms and the bunch width (full width at half-maximum) was set to $\sim$1 $\mu$s without degrading the typical resolution of $\sim$80~MHz of the hf spectrum. A Sirah Matisse TS Ti:sapphire ring laser was used to produce 787 nm light that was subsequently frequency doubled to 393~nm light using a SpectraPhysics Wavetrain. 
The Matisse was stabilized using a HighFinesse WSU-30 wavelength meter, calibrated with a frequency-stabilized He-Ne laser. The laser-light power was stabilized at 300~$\mu$W, which gave a maximum signal to noise ratio, using a laser power controller \cite{beoc}.
Two identical photon detectors were used in series along the beam line to detect the resonant fluorescence. 
A scanning voltage was applied to the light collection section to vary the incoming ion beam velocity so that the Doppler-shifted laser frequency could be tuned into resonance with the hf transitions. 

\begin{figure}[b]
\includegraphics[width=1\linewidth]{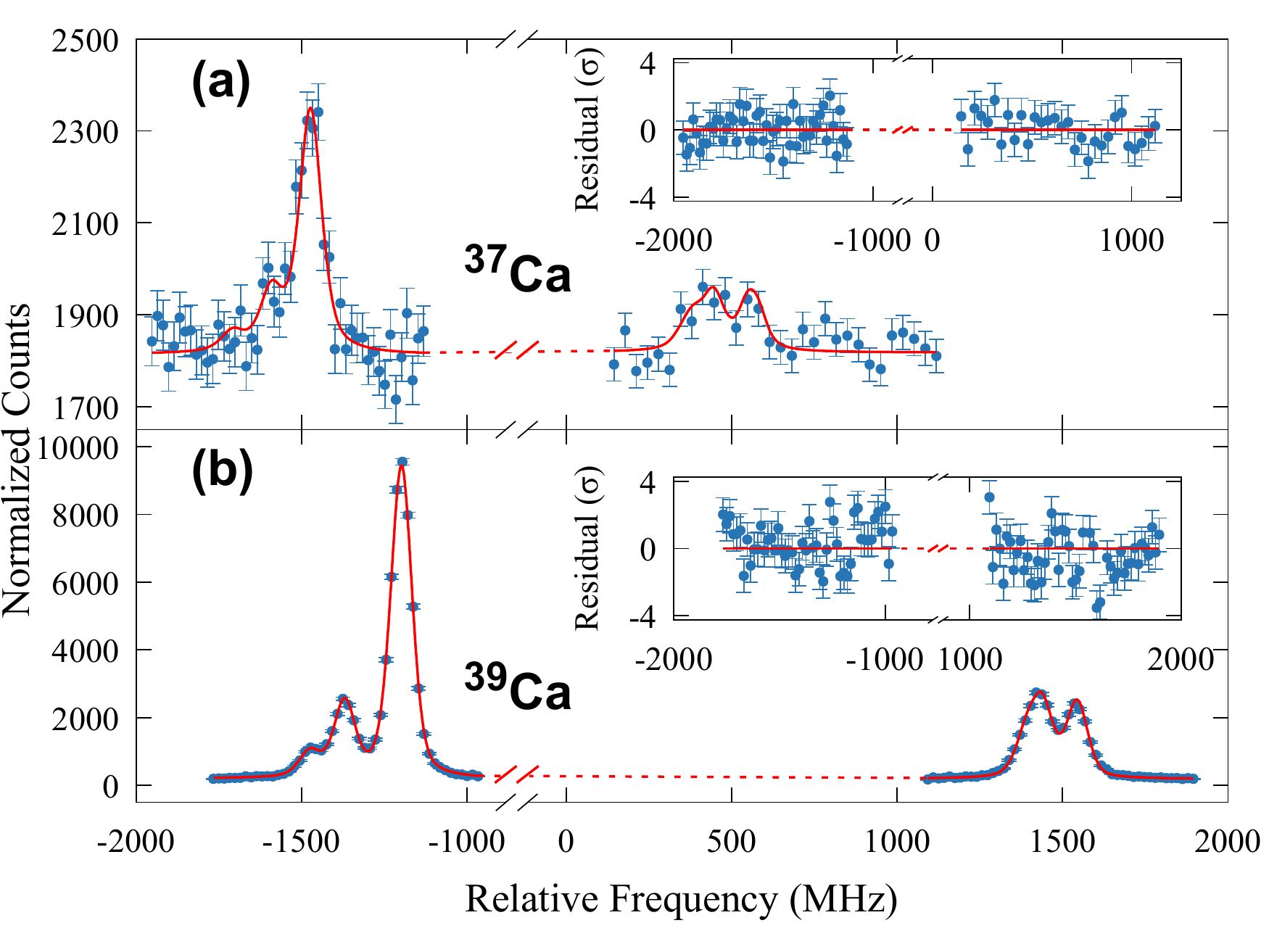}
\caption{Hyperfine spectra and residuals of (a) $^{37}$Ca and (b) $^{39}$Ca. The solid circles are the data and the solid line is the best fit of a Voigt profile.}
\label{fig:37Ca_hfs}
\end{figure}

{\it Experimental results} ---
The obtained hf spectrum of $^{37}$Ca is shown in Figure \ref{fig:37Ca_hfs}. 
The hf spectrum of $^{39}$Ca(3/2$^+$, 859.6 ms) was also measured in the present study. The $^{39}$Ca beam was produced in a similar procedure as $^{37}$Ca, and the hf spectrum was measured with the same laser power as $^{37}$Ca to be used as a line shape reference in the fitting of $^{37}$Ca. There are six allowed hf transitions between the $^2S_{1/2}$ and $^2P_{3/2}$ states since the nuclear spins of $^{37, 39}$Ca are $I$ = 3/2. The shift of a hf level is given by 
\begin{equation}
\Delta E = \frac{K}{2}A^{\rm hf} + \frac{3K(K+1) - 4I(I+1)J(J+1)}{8I(2I-1)J(2J-1)}B^{\rm hf}
\end{equation}
where $A^{\rm hf}$ and $B^{\rm hf}$ are the magnetic and quadrupole hf coupling constants, respectively, $K = F(F + 1)-I(I + 1)-J(J + 1)$, $I$ is the nuclear spin, $J$ is the total electronic angular momentum and $\bm{F = I + J}$. The hf coupling constants are defined as
$A^{\rm hf} = \mu B_0/IJ$ and $B^{\rm hf} = eQV_{zz}$. Here, $\mu$ and $Q$ are the magnetic-dipole and spectroscopic electric-quadrupole moments of the nucleus, respectively, $B_0$ and $V_{zz}$ are the magnetic field and the electric field gradient, respectively, generated by orbital electrons at the position of the nucleus and $e$ is the elementary charge. 
The $B_0$ and $V_{zz}$ are isotope-independent, assuming a point-like nucleus. 

A Voigt profile \cite{sta08} was used in the fitting of the hf spectrum. All six peaks of the $^{39}$Ca hf spectrum were fitted with a common line shape and width, and free parameters of ground and excited states hf coupling constants, Lorentz fraction and intensities of each peak. The high ion-beam rate allowed the reliable determination of these parameters with high precision. The $^{37}$Ca hf spectrum was fitted with the relative peak intensities and Lorentz fraction constrained to those determined in the $^{39}$Ca analysis.  Also the ratio between $A^{\rm hf}(^2S_{1/2})$ and $A^{\rm hf}(^2P_{3/2})$ was fixed to 26.24(4) deduced from the $^{39}$Ca fit. The obtained hf coupling constants are summarized in Table \ref{table:hfc}. It is noted that the $A^{\rm hf}$-factor ratio was determined in the previous measurement as 25.92(3) \cite{gar15} and deviates from the present ratio. The reason is not known. Variation in fitted hyperfine coupling constant due to the deviation of $A^{\rm hf}$-factor ratios was taken into account as systematic uncertainties. It is also noted that the $A^{\rm hf}$ and $B^{\rm hf}$ parameters are highly correlated in the fit. The statistical error for the results of $^{37}$Ca in Table \ref{table:hfc} takes the correlation into account to be conservative, and used in the present analysis. Without the correlation, statistical errors of the $A^{\rm hf}(^2S_{1/2})$ and $B^{\rm hf}(^2P_{3/2})$ of $^{37}$Ca are 6.3 MHz and 8.4 MHz, respectively. The quadratic sum of statistical and systematical uncertainty was taken as total uncertainty in the following discussion.

Unknown nuclear moments may be deduced from hf coupling constants using a reference nucleus of the same element, whose hf coupling constants for the same electronic level, nuclear spin and electromagnetic moments are known. 
A simple ratio of hf coupling constants derives nuclear moments as $\mu = \mu_{\rm R}\frac{A^{\rm hf}}{A^{\rm hf}_{\rm R}}\frac{I}{I_{\rm R}}$ and $Q = Q_{\rm R}\frac{B^{\rm hf}}{B^{\rm hf}_{\rm R}}$, where the subscript R indicates a reference nucleus. 
In the present study, $^{43}$Ca($I$ = 7/2) was employed as a reference, and $A^{\rm hf}(^2S_{1/2})$ = $-$806.40207160(8) MHz \cite{arb94} and $\mu$ = $-$1.317643(7)$\mu_N$ \cite{lut73} were used to extract the magnetic moment. 
A theoretical value of $eV_{zz}$ = 1.513(7) MHz/fm$^2$ \cite{sah09} was used for the extraction of $Q$ since a sufficiently precise measurement of $B^{\rm hf}_{\rm R}(^2P_{3/2})$ does not exist. 

\begin{table}[b]
	\caption{\label{table:hfc}
The obtained hf coupling constants of $^{37, 39}$Ca for the $4s$~$^2S_{1/2}$ and $4s$~$^2P_{3/2}$ states. The first and second parentheses contain uncertainties due to statistical and systematic errors, respectively. The systematic errors are from high voltage calibrations and the variation of the $A^{\rm hf}$-factor ratio from the literature value, which dominates the systematic error.}
	\begin{ruledtabular}
	\begin{tabular}{lcccc}
		&& \multicolumn{2}{c}{$A^{\rm hf}$ (MHz)} & $B^{\rm hf}$ (MHz)\\
		\cline{3-4}\cline{5-5}
		$A$ & $I^\pi$ & $^2S_{1/2}$ & $^2P_{3/2}$ & $^2P_{3/2}$\\   
		\hline
		$37$ & 3/2$^+$ & +1064.5(103)(08) & +40.57(39)(27) & $-$22.9(163)(05)\\
		$39$ & 3/2$^+$ & +1457.20(14)(34) & +55.53(9)(32) & +5.79(26)(32)
	\end{tabular}
	\end{ruledtabular}
\end{table}
%
\begin{table*}[t]
	\caption{\label{table:moments}
Results for the magnetic moments of 3/2$^+$ states for $Z = 20$ ($^{39}$Ca and $^{37}$Ca) and $N = 20$ ($^{39}$K and $^{37}$Cl). 
Numbers are given in the unit of $\mu_N$ except $\left<s_z\right>$. 
The effective $g$-factors are taken from the Table I of \cite{ric08} for six-parameter form of the M1 operator. 
The $\mu$(IS) and $\mu$(IV) are defined as $\mu$(IS/IV) = $\mu$($T_3 = +T$) $\pm$ $\mu$($T_3 = -T$).
}
	\begin{ruledtabular}
	\begin{tabular}{lcccccc}
	$A$ && $Z = 20$ & $N = 20$ & $\mu$(IS) & $\left<s_z\right>$ & $\mu$(IV)\\
	\hline
	& sp $g^{\rm free}$ & +1.148 & +0.124 & +1.272 & $-$0.600 & +1.024\\
	\hline
	39 & Exp. & +1.0217(1) \cite{min76} & +0.3915073(1) \cite{sah74} & +1.4131(1) & $-$0.2284(3) & +0.6302(1)\\
	& sp $g^{\rm eff}$ & +0.930 & +0.469 & +1.399 & $-$0.266 & +0.461\\
	& VS-IMSRG & +1.349 & $-$0.035 & +1.314 & $-$0.488 & +1.384\\
	\hline
	37 & Exp. & +0.7453(72) & +0.6841236(4)  \cite{bla72} & +1.429(7) & $-$0.19(2) & +0.061(7)\\
	& USDA-EM1 & +0.770 & +0.677 & +1.447 & $-$0.139 & +0.093\\
	& USDB-EM1 & +0.754 & +0.675 & +1.429 & $-$0.187 & +0.079\\	
	& VS-IMSRG & +1.055 & +0.290 & +1.345 & $-$0.409 & +0.765\\
	\end{tabular}
	\end{ruledtabular}
\end{table*}

The spectroscopic quadrupole moments were determined to be $Q(^{37}{\rm Ca}) = -15(11)~e^2{\rm fm}^2$ and $Q(^{39}{\rm Ca}) = +3.82(27) ~e^2{\rm fm}^2$.
The present $Q$($^{39}$Ca) is consistent with the previous value \cite{mat99} determined using the $\beta$-NMR technique, and three times more precise than the previous value with experimental determination of the prolate deformation (the positive sign).  The $Q$($^{37}$Ca) is determined for the first time in the present study including its oblate deformation (the negative sign). In the shell model calculation with the USDB interaction, discussed later in this Letter, effective charges, $e_{\rm p}$ = 1.5 and $e_{\rm n}$ = 0.5, gives $Q$($^{37}$Ca) = $-$2.6 $e^2$fm$^2$. The agreement is fair, but no further discussion is made here because the present value has a large uncertainty due to the low signal-to-noise ratio to resolve the $^2P_{3/2}$ splitting in the hf spectrum. 

The magnetic moment of $^{37}$Ca was determined from the $A^{\rm hf}$($^2S_{1/2}$) to be $\mu(^{37}{\rm Ca}) = +0.7453(72)~\mu_N$.
The result is summarized in Table \ref{table:moments}.
It is noted that the hf anomaly is neglected in the extraction of the magnetic moment. The hf anomaly, $^1\Delta^2$, is caused by the difference of the nuclear magnetization distribution \cite{boh50} between two isotopes 1 and 2, and is given by $A^{\rm hf}_1/A^{\rm hf}_2 \approx g_1/g_2\left(1 + ^1\Delta^2\right)$, where the $g$ factor is defined as $g = \mu/I$. 
For $^{39, 43}$Ca, there exist independent measurements of $A^{\rm hf}$($^{43}$Ca) \cite{arb94}, $g$($^{43}$Ca) \cite{lut73} and $g$($^{39}$Ca) \cite{min76}. The hf anomaly can be deduced together with the present value of $A^{\rm hf}$($^{39}$Ca) for the $^2S_{1/2}$ state as $^{43}\Delta^{39} = +0.0012(3)$. The hf anomaly between $^{37}$Ca and $^{43}$Ca is expected to be similar to $^{43}\Delta^{39}$, and the contribution to $\mu$($^{37}$Ca) is negligible compared to the experimental uncertainty.

{\it Discussion} --- 
The magnetic moments of $^{37}$Ca and $^{37}$Cl with one particle in the $d_{3/2}$ shell, and their counterparts $^{39}$Ca and $^{39}$K with one hole in the $d_{3/2}$ shell provide a unique opportunity to study the first-order core-polarization model for the nucleon configuration mixing \cite{ari54a,ari54b}. 
The first-order corrections are important for closed-shell nuclei, where the $j_{>} = l +1/2$ component of the spin-orbit pair is mostly filled, and the $ j_{<} = l - 1/2 $ component is mostly empty, which we will call the $  JJ  $ closed-shell configuration.
The $^{36}$Ca wavefunction is dominated by the $  (d_{5/2})^{6} (s_{1/2})^{2}  $ $  JJ  $-type configuration for neutrons, where the $  d_{5/2}  $ orbital is filled and the $  d_{3/2}  $
orbital is empty. 
When both $  j_{>}  $ and $  j_{<}  $ orbitals are mostly filled, which we will call the $  LS  $ closed-shell configuration, the core-polarization effect is small. 
The $^{40}$Ca wavefunction is dominated by the $  (d_{5/2})^{6} (s_{1/2})^{2} (d_{3/2})^{4}  $~$  LS  $-type configuration for neutrons.
It is noted that both $^{36}$Ca and $^{40}$Ca have an $LS$ closed shell for protons. 
The observable associated with this change from $JJ$ to $LS$ closed shell configurations is the $\mu$($^{37}$Ca), which has one particle outside of the $^{36}$Ca neutron $  JJ  $ core, relative to that of $^{39}$Ca with one hole inside the $^{40}$Ca neutron $  LS  $ core.
We can also observe the similar transition in their mirror $\mu$($^{37}$Cl) (one particle outside of the $^{36}$S proton $  JJ  $ core) relative to that of $^{39}$K (one hole inside the $^{40}$Ca proton $  LS  $ core).

For the calculations we use the $sd$-shell model space with the USDA and USDB Hamiltonians \cite{bro06}. 
The magnetic moment (M1) operator is defined as ${\bm \mu} = g_l\left<{\bm l}\right> + g_s\left<{\bm s}\right> + g_p\left<[{\bm Y_2}, {\bm s}]\right>$, where $l$, $s$, $p$ represent the orbital angular momentum, spin and tensor terms, respectively.
The results of the calculations are summarized in Table~\ref{table:moments}.
The free nucleon $g$ factors ($g_l^{\rm p}$ = 1, $g_l^{\rm n}$ = 0, $g_s^{\rm p}$ = 5.586, $g_s^{\rm n}$ = $-$3.826) were used for single-particle (Schmidt) values denoted as ``sp $g^{\rm free}$". All other calculations were performed with the effective $g$ factors that are obtained from a six-parameter fit to other magnetic moments in the $A$ = 16 - 40 mass region \cite{ric08}.
The results labeled USDA-EM1 and USDB-EM1 are given in Table~\ref{table:moments} and discussed in this paper.


The first-order core polarization is contained within the $sd$-shell model space for present calculations, and the effective $g$-factors reflect higher-order corrections due to correlations beyond the $  sd  $ model space and meson-exchange currents \cite{ric08}.
For the $  d_{3/2}  $ orbital the effective single-particle magnetic moment for $A$ = 37 or single-hole magnetic moment for $A$ = 39 denoted as ``sp $g^{\rm eff}$" for neutrons (protons) is 0.930$\mu_N$ (0.469$\mu_N$) compared to the single-particle value of 1.148$\mu_N$ (0.124$\mu_N$), and the variation indicates the contribution from the higher-order corrections through $g^{\rm eff}$.
\begin{table}[t]
	\caption{\label{table:theory_1}
Wave function for $^{37}$Ca(3/2$^+$). The occupation for each configuration is shown in \%.}
	\begin{ruledtabular}
	\begin{tabular}{lccc}
		($n$5, $n$1, $n$3) & USDA-EM1 & USDB-EM1 & VS-IMSRG\\
		\hline
		(621) & 94.61 &  95.03 & 90.28\\
		(603)   & 1.68 & 1.64 & 2.81\\
		(612)  & 0.32 &  0.68 & 0.51\\
		(423)  & 2.63 &  2.21 & 5.61\\
		(513)  & 0.51 &  0.23 & 0.29\\
		(522)  & 0.23 &  0.20 & 0.45\\
		\hline
		$\mu$ ($\mu_N$) & 0.770 & 0.754 & 1.055		
	\end{tabular}
	\end{ruledtabular}
\end{table}

The experimental magnetic moment for $^{37}$Ca is in excellent agreement with USDA/B-EM1 calculations.
The wavefunctions for $^{37}$Ca are given in Table~\ref{table:theory_1} in terms of the percent probabilities for the six allowed partitions. 
The partitions are given in terms of the number of neutrons that occupy each orbital ($n$5, $n$1, $n$3) as in $(d_{5/2})^{n5} (s_{1/2})^{n1} (d_{3/2})^{n3} $.
The partitions that are important for the magnetic moment are (621), (603) and (522).
The magnetic moments for the (621) and (603) partitions are just the single-particle value of 0.930$\mu_N$. 
The interference of the (621) and (522) partitions decreases the magnetic moment to 0.760$\mu_N$/0.750$\mu_N$ for USDA/B-EM1. 
The addition of the other four partitions give a final result of 0.770$\mu_N$/0.754$\mu_N$ for USDA/B-EM1.
These are in excellent agreement with experiment.

The mixing of the (522) partition with the  $JJ$ closed-shell partition (620) plus one neutron particle in the $d_{3/2}$, (621), is the core-polarization effect, where one neutron is
moved from the $d_{5/2}$ to the $d_{3/2}$ orbital \cite{ari54a,ari54b}. 
On the other hand, the $LS$ closed-shell partition (604) plus one neutron hole in the $d_{3/2}$, (603), has no core-polarization correction. 
This is also true for the $LS$ closed shell partition (624) plus one neutron hole in $d_{3/2}$ (623) that is the $sd$-shell configuration for $^{39}$Ca. 
Thus, the magnetic moment is sensitive to the mixing between the $s_{1/2}$ and $d_{3/2}$ orbitals, via the relative amounts of the (621) and (603) partitions. 
The agreement with experiment indicates that $^{37}$Ca is dominated by the (621) partition that represents $^{36}$Ca in a  $(d_{5/2})^6 (s_{1/2})^2$ closed sub-shell configuration plus one neutron in the $d_{3/2}$ orbital.   

The same situation occurs for the magnetic moment of the mirror nucleus $^{37}$Cl. 
The effective single-particle value of 0.469$\mu_N$ is increased by the core-polarization to 0.675$\mu_N$, which is also in good agreement with the experimental value.

\begin{figure}[b]
\includegraphics[width=1\linewidth]{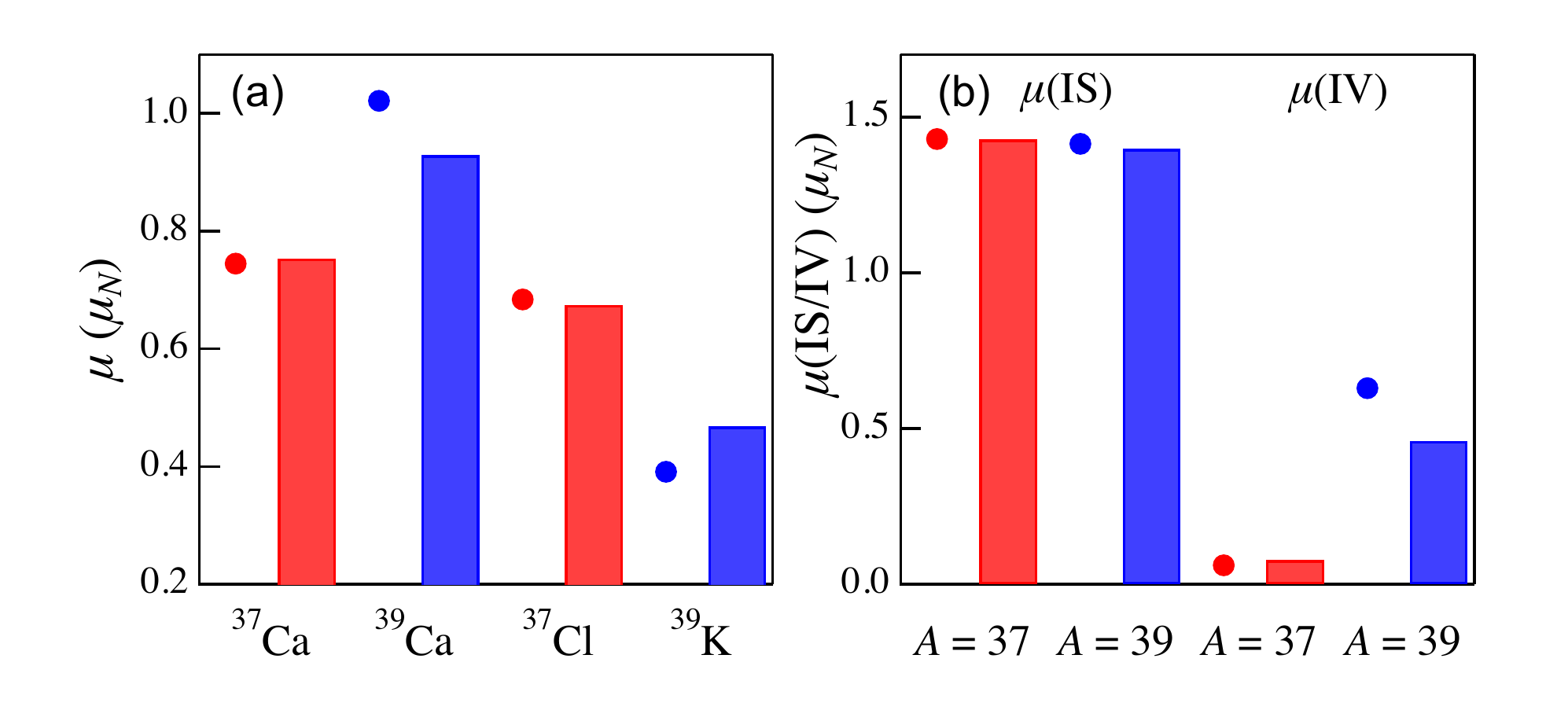}
\caption{Comparison between experimental values (solid circles) and shell-model calculations with USDB-EM1 (bars) for (a) magnetic moments and (b) isoscalar and isovector magnetic moment.}
\label{fig:exp_theo_comp}
\end{figure}
As the configuration is varied from the $JJ$ type for $^{37}$Ca to the $LS$ type for $^{39}$Ca, the magnetic moment is expected to increase from 0.754$\mu_N$ (USDB) to 0.930$\mu_N$ (sp $g^{\rm eff}$).
The experimental values increase from 0.7453(72)$\mu_N$ to 1.0217(1)$\mu_N$ \cite{min76}. 
The agreement of $^{37}$Ca is excellent but the calculation for $^{39}$Ca underestimates the experimental value as shown in Figure \ref{fig:exp_theo_comp}.
This increase can be understood by observing that the ground state $I$ = 3/2 of $^{37}$Ca can be obtained with the excitation of neutrons from the $d_{5/2}$ to the $d_{3/2}$ shell, whereas for the $I$ = 3/2 state of $^{39}$Ca such excitation is prohibited in the $sd$-shell model space. 
A similar behavior can be seen in their mirror magnetic moments. Going from $^{37}$Cl to $^{39}$K we expect a reduction of the magnetic moment from 0.675$\mu_N$ to 0.469$\mu_N$, whereas the experimental values decrease from 0.6841236(4)$\mu_N$ \cite{bla72} to 0.3915073(1)$\mu_N$ \cite{sah74}.

The isoscalar $\mu$(IS) and isovector $\mu$(IV) parts of a magnetic moment are also evaluated, which are deduced as $\mu$(IS/IV) = $\mu$($T_3 = +T$) $\pm$ $\mu$($T_3 = -T$) with the isospin $T_3$ = +1/2 for protons.
The isoscalar spin expectation value $\left<s_z\right>$ was evaluated as $\mu$(IS) = $I$ + 0.38$\left<s_z\right>$ \cite{sac46, sug69} and also listed in Table \ref{table:moments}.
The change in $\mu$(IS) from the single-particle value of 1.272 to those deduced from experimental moments of 1.4131(1) for $A = 39$ and 1.429(7) for $A = 37$, is mainly due to the change of $g^{\rm free}$ to $g^{\rm eff}$. 
This can be seen in Figure \ref{fig:exp_theo_comp} as the ``sp $g^{\rm eff}$" value (blue bar) for $\mu$(IS) well explains the experimental values for $A$ = 37 and 39.
On the other hand, there is a large reduction in $\mu$(IV) coming both from the change of $g^{\rm free}$ to $g^{\rm eff}$ and from the core-polarization within the $sd$-model space.
The $\mu$(IV) is reduced to be near zero for $A = 37$ and the shell model calculation reproduces the experiment, however for $A = 39$ the discrepancy is large.  
The good agreement for $^{37}$Ca and $^{37}$Cl confirm the importance of core-polarization for the $  JJ  $ closed-shell nuclei.
The larger deviation between theory and experiment for $^{39}$Ca and $^{39}$K than those for $^{37}$Ca and $^{37}$Cl indicates that additional non-$sd$-shell components of $^{40}$Ca are larger than those of $^{36}$Ca and $^{36}$S. 
It is also noted that magnetic moments of heavier $^{41,43,45}$Ca suggest large nucleon excitations across the $sd$ shell around the neutron number $N$ = 20 \cite{gar15}.
In this regard it appears that the $^{36}$Ca ($^{36}$S) nucleus may be a better closed sub-shell nucleus at $N$ = 16 ($Z$ = 16) than the $^{40}$Ca nucleus.

We also calculate magnetic moments of $A$ = 37 and 39 pairs using the valence-space formulation of the ab initio in-medium similarity renormalization group (VS-IMSRG) \cite{tsu12,Bogn14,Stro16,Stro17}. In this approach we consistently transform the $M1$ operator and no effective $g$ factors were used \cite{Parz17,Hend18}. The 1.8/2.0 chiral interaction defined in Refs.~\cite{Hebe11,Simo16,Simo17} was taken as the initial two- and three-nucleon potentials within a harmonic oscillator basis of 13 major shells, a frequency $\hbar \omega = 16$ MeV, operators truncated at the two-body level and $sd$-shell model space with a $^{16}$O core. 
The results are summarized in Table~\ref{table:moments} and \ref{table:theory_1}.
The calculations for $^{37,39}$Ca overestimate the experimental values, but have the canonical nucleon configuration of 90\% and confirm a closed sub-shell structure of $^{36}$Ca. 
Compared to the USDA/B-EM1 calculations, the VS-IMSRG agrees with the dominance of (620) partition for $^{36}$Ca. 
However, the amount of the (522) partition that gives the core-polarization correction is a factor of two larger. The deviation is likely due to meson-exchange currents \cite{bac14}, which are not included in the present VS-IMSRG calculations, but are included indirectly through the effective $g$ factors in the USDA/B-EM1 calculations. 

{\it Summary} --
Bunched-beam collinear laser spectroscopy was performed to determine electromagnetic moments of $^{37}$Ca to probe the closed sub-shell nature of the $^{36}$Ca nucleus.
Shell-model calculations were performed in the $sd$-shell-model space with the USDA/B-EM1 interaction and effective $g$ factors. 
The calculated $\mu$($^{37}$Ca) reproduces the experimental value with a 95\% probability of the canonical nucleon configuration, yet the first-order core polarization effect within the $sd$ shell is critical for the agreement.  
The calculated value for $\mu$($^{39}$Ca), which has one neutron hole inside the $^{40}$Ca core, shows poor agreement with the experimental value compared to that of $\mu$($^{37}$Ca).
A similar behavior can be seen in magnetic moments of the mirror partners $^{37}$Cl and $^{39}$K.
This indicates that additional non $sd$-shell components of $^{40}$Ca are larger than those of $^{36}$Ca. 
The $^{36}$Ca nucleus appears to be a better closed sub-shell nucleus at $N$ = 16 than $^{40}$Ca as represented by the USDA/B-EM1 Hamiltonian.
The ab-initio VS-IMSRG calculations give reasonable agreement with experimental $\mu$($^{39, 37}$Ca) and confirms a closed sub-shell structure of $^{36}$Ca. 

\begin{acknowledgments}
This work was supported in part by the National Science Foundation, grant Nos.~PHY-15-65546 and PHY-18-11855; the U. S. Department of Energy, National Nuclear Security Administration, Grant No.~DE-NA0002924; the U.S. Department of Energy, Office of Science, Office of Nuclear Physics, Grant No.~DE-AC05-00OR22725 with UT-Battelle, LLC; Natural Sciences and Engineering Research Council of Canada under grant number SAPPJ-2017-00039; and the Deutsche Forschungsgemeinschaft (DFG, German Research Foundation) -- Projektnummer 279384907 -- SFB 1245.
\end{acknowledgments}

\bibliographystyle{apsrev4-1} 
\bibliography{becolabib_v1}

\end{document}